\documentstyle[prl,aps]{revtex}
\draft
\begin{document}
\title{Master Equations for Extremal Models}
\author{Daniele Balboni}
\address{D\'epartement de Physique Th\'eorique, Universit\'e de Gen\`eve,
CH-1211 Gen\`eve 4, Switzerland.}

\maketitle
\begin{abstract}
A general method to derive the master equations for extremal models is established.
These systems are shown to develop a peculiar kind of correlations between
elements related to the characterization of extremal dynamics as an information 
process.

\vspace {0.3truecm} 
{PACS numbers: 87.10.+e, 05.40+j}
\end{abstract}

\section{Introduction}

In this article we would like to establish the general method and 
the proper way to derive the master equations
governing the Bak-Sneppen type models of extremal dynamics
\cite{BakSne,PacMasBak}.
Apart from the relevance as rough models of biological evolution, the interest
of these systems is that they show at equilibrium a self organized critical 
state, with power law
correlations in both time and space (if any 'space' relation between elements
is defined, as in the case of the nearest neighbor model) 
and the phenomena of avalanches and
punctuated equilibrium. 
The importance of the extremal dynamics in determining the self organized critical properties
of several models (another example is the Invasion Percolation) 
has been discussed in \cite{GabCafMarPie}.
 
The systems under consideration are composed of a set of, say, $L$ elements 
characterized by the value of 
a variable, usually a one dimensional variable taken in the unit interval
(in the evolutionary interpretation it parametrizes the facility to mutate). 
Let's denote it $x_A$ where $A=1,\dots,L$ is the element label 
and $x_A\in[0,1]$.
The characteristic feature of the dynamics of these systems is the 
selection, at each update, 
of the element which has the maximum value (the "maximum element") and the 
reassignment to it of a new value randomly chosen in the unit interval.
This kind of rule is referred to as extremal dynamics. 
In addition new random values are assigned to a certain number
of other elements which, in the more interesting case, have a neighborhood 
relation with the maximum element (nearest neighbor models, 
n.n. models in the following). 
Another possibility is to choose them at random
(random models), in which case the model is analytically solvable
via a master equation of the diffusion type
for the probability $P_\lambda(n)$ to have
$n$ elements with values less than a given threshold $\lambda$
\cite{deBDerFlyJacWet}.
Whereas this latter model has the advantage of
giving exact results, it is not very rich, lacking any spatial structure
and in some sense not being an interacting system, as the maximum element 
affects the other elements just by triggering the change of their values.
To write a master equation for the n.n. model it is not anymore 
possible to work
with the statistics of the variables $x$ disregarding the precise assignments
to the elements, as it is the case in the random model.
Instead we have to cope with the full probability density $p(\vec{x})$, 
$\vec{x}=(x_1,\dots,x_L)$, that is with probabilities defined in
the hypercube $[0,1]^L$. To our knowledge the master equation for the n.n.
model has not yet been derived.

Now an interesting point is the characterization of extremal dynamics as an
information process: at each time step the choice of the maximum element
takes into account the states of all the elements 
(or at least a large
number of them: in Invasion Percolation the perimeter elements) giving information about them.
As we'll see below, this global 
information creates correlations even between elements that remain inactive.
We'll refer to them as information correlations.
As pointed out in \cite{deBJacWet} the role of fluctuations is essential 
to determine the properties of the  
self organized critical state of the Bak-Sneppen models, 
in particular the avalanche phenomenon
and the distribution of the maximum element: in fact the relevant quantity,
the number of active elements above self organized threshold, 
is of the same order of magnitude as its fluctuations. 
This limits severely a mean field theory treatment where correlations between
elements are neglected, and that even in the case of the random model 
where they arise only as information correlations.

\section{How correlations between elements are generated}

The key feature of our method is to decompose the hypercube in a partition
where the regions correspond to a given ordering of the element values.
Let's start by introducing some definitions and notations.
$I_{A_1,\dots,A_n}$ will denote the domain of the hypercube where the values of
the $n$ elements $A_1,\dots,A_n$ are decreasingly ordered:
\begin{equation}
I_{A_1,\dots,A_n}\equiv\{\vec{x}\in[0,1]^L~/~
x_{A_1}\ge\dots\ge x_{A_n}\ge x_B~~~\forall B\not=A_1,\dots,A_n\}
\label{equ:1}\end{equation}
and $\Delta_{A_1,\dots,A_n}(\vec{x})$ will denote the characteristic function 
which has this domain as support:
$$\Delta_{A_1,\dots,A_n}(\vec{x})=
\cases{1,&if $\vec{x}\in I_{A_1,\dots,A_n}$;\cr 0, &otherwise.\cr}$$
First we have to determine how the choice of the maximum element 
as the active one
affects the probability distribution. 
If at a given time step the probability distribution is $p(\vec{x})$, 
the knowledge that the maximum element is the element $T$  
changes $p$ in the following way:
$$p(\vec{x})\rightarrow p(\vec{x}|I_T)=N\Delta_T(\vec{x})p(\vec{x}).$$
The normalization factor $N$, which depends in general on $T$, is the inverse 
of the probability to have $T$ as the maximum element and $N=L$ 
if the function $p$ is symmetric.
If we now replace its value with a new one randomly chosen in the unit interval
and with probability distribution $\varphi(x)$, the updated distribution
function becomes:
\begin{equation} 
p(\vec{x})\rightarrow p(\vec{x};T)=\varphi(x_T)\int_0^1 dx_T~p(\vec{x}|I_T)
=N\varphi(x_T)\int^1_{\rm{max}\{x_A;A\not=T\}}dx_T~p(\vec{x}).
\label{equ:10}
\end{equation}
In the following we'll consider the case $\varphi(x)=1$.
Now we want to show that this new distribution represents correlated 
elements even if $p(\vec{x})$ does not, that is the
knowledge of which is the maximum element creates correlations between the other
elements.  
To see this let's consider $L=3$ and $p(x,y,z)=1$. If we 
are now given the information that the first element, corresponding to $x$, is
the biggest one, then the probability distribution restricted to $y$ and $z$ is:
$$\rho(y,z)=\int_0^1 dx~p(x,y,z|I_1)= 
N\left[\theta(y-z)(1-y)+\theta(z-y)(1-z)\right]$$
and
$$\rho(z)=\int_0^1 dy~\rho(y,z)=N{{(1-z^2)}\over{2}}.$$
Then the conditional probability $\rho(y|z)=\rho(y,z)/\rho(z)$
depends on $z$, that is $y$ and $z$ are no more independent: an information 
on one of the two, let's say $z$, gives information on the other. 
To understand this suppose $z=\xi$; this means $x>\xi$
and, because $y<x$, we have $\rho(y|z=\xi)={\rm const.}$ 
for $y<\xi$. On the other hand $\rho(y|z=\xi)$ has to tend to $0$ for 
$y\rightarrow 1$: if $y=1-\epsilon$ then $x>1-\epsilon$ and the probability
of this last condition tends to zero for $\epsilon\rightarrow0$.
This result contrasts with what is said in \cite{Mar1,Mar2,GabCafMarPie}, 
where it is erroneously assumed
that the element variables are independent after the application of the 
extremal rule, and that the probability distribution 
maintains a factorized form.
As we already pointed out the origin of these correlations comes from
an information process and not from an action connecting the elements.
This last mechanism is only present in the n.n. model where it is responsible of spatial 
correlations. It has been shown in \cite{deBDerFlyJacWet} 
that in the random model mean field and exact
results (supplied by simulations) have strong discrepancies and we see now that
they can be attributed entirely to these information correlations.

\section{The simplest model}

To show the technique used to derive and cast in a convenient form
the master equations of the Bak-Sneppen models,
we examine in this section the quite simple case where the maximum 
element is the only one to be updated. 
The equation we are looking for is that which gives the 
probability distribution at time step
$i+1$, $p_{i+1}(\vec{x})$, in terms of that at time $i$, $p_i(\vec{x})$.
If $\mu_i(T)$ is the probability to have element $T$ as the 
maximum at time $i$ and noting that it is 
the inverse of the normalization factor $N$ in equation (\ref{equ:10}), the
master equation for the simplest model is:
$$p_{i+1}(\vec{x})=\sum_{T=1}^L\mu_i(T) p_i(\vec{x};T)
=\sum_T\int^1_{\rm{max}\{x_A;A\not=T\}}dx_T~p_i(\vec{x}).$$
Our strategy in investigating this equation is now to decompose the hypercube
$[0,1]^L$ in regions of the type (\ref{equ:1})
such that the probability distribution maintains an analytic form 
in each cell. Of course the finest partition where we take as cells 
the regions $I_{A_1,\dots,A_L}$ where $L$ is the total number of elements 
does the job for every model. In the very simple
case under consideration it turns however out that the partition in cells
$I_{A,B}$ is sufficient. 
Then we decompose the probability distribution in the following way:
$$p(\vec{x})=\sum_{A,B}\Delta_{A,B}(\vec{x})p^{(A,B)}(\vec{x})$$
where the functions $p^{(A,B)}$ can be extracted from $p$ by using $\Delta_{A,B}$
as a projector
$$p^{(A,B)}=\Delta_{A,B}~p$$
and we want to express the functions $p_{i+1}^{(A,B)}$ in terms of the 
$p_i^{(C,D)}$'s.  
It turns out that they remain analytic after the probability transition 
$i\rightarrow i+1$. To manipulate these "pieces" in the master equation where integrals
are taken over domains of integration that cross the boundaries of 
the partition, we need to
take into consideration other kinds of regions:
$$I_{A_1,\dots,A_n}^D
\equiv\{\vec{x}\in[0,1]^L~/~
x_{A_1}\ge\dots\ge x_{A_n}\ge x_B~~~\forall B\not=D,A_1,\dots,A_n\}$$
and, as before:
$$\Delta_{A_1,\dots,A_n}^D(\vec{x})=
\cases{1,&if $\vec{x}\in I_{A_1,\dots,A_n}^D$\cr 0, &otherwise.\cr}$$
By means of the rules:
$$\Delta_T\Delta_{A,B}=\delta_{T,A}\Delta_{A,B}$$
$$\Delta_{D,E}\Delta^T_B=\Delta_{D,E}(\delta_{B,D}+\delta_{B,E}\delta_{T,D})$$
we have, for a symmetric function $p(\vec{x})$ and $N=L$:
$$p(\vec{x}|I_T)=L\Delta_T(\vec{x})p(\vec{x})
=L\sum_{A,B}\Delta_T\Delta_{A,B}p^{(A,B)}
=L\sum_B\Delta_{T,B}p^{(T,B)}$$
$$p(\vec{x};T)
=\int dx_T~p(\vec{x}|I_T)
=L\sum_B\Delta^T_B\int^1_{x_B}dx_T~p^{(T,B)}$$
then 
$$p_{i+1}(\vec{x})={1\over L}\sum_T p_i(\vec{x};T)
=\sum_{T,B}\Delta^T_B\int^1_{x_B}dx_T~p^{(T,B)}$$
and finally
$$p_{i+1}^{(D,E)}=\Delta_{D,E}~p_{i+1}
=\Delta_{D,E}\left[\sum_{T(\not=D)}\int^1_{x_D}dx_T~p_i^{(T,D)}
+\int^1_{x_E}dx_D~p_i^{(D,E)}\right]$$
which is the desired result, that is an analytic form for $p^{(D,E)}$
in its domain of definition $I_{D,E}$.
If $p_0=1$ this recursion equation is easily solved to give:
$$p_i^{(D,E)}(\vec{x})
={{(i-1+L)!}\over{L!i!}}\left[(L-1)(1-x_D)^i+(1-x_E)^i\right].$$
The function $p_i(\vec{x})$ encodes clearly the most detailed
information on the system at time $i$. We can calculate from it
for example 
the one-element probability distribution. Choosing the element $1$:
$$\rho_i(x_1)=\int_0^1dx_2\int_0^1dx_3\cdots\int_0^1dx_L~p_i(\vec{x})$$ 
$$=(L-1)\int_0^{x_1}dx_2\int_0^{x_2}dx_3\cdots\int_0^{x_2}dx_L~p_i^{(1,2)}$$
$$+(L-1)\int_{x_1}^1dx_2\int_0^{x_1}dx_3\cdots\int_0^{x_1}dx_L~p_i^{(2,1)}$$
$$+(L-1)(L-2)\int_0^1dx_2\int_{x_1}^{x_2}dx_3\int_0^{x_3}dx_4
\cdots\int_0^{x_3}dx_L~p_i^{(2,3)}.$$
The first term is the probability distribution (to be normalized) of the 
maximum element, the second is that of the next smaller one, 
and the third term the
probability distribution of any of the others (in this case they are
all equally distributed). 
By substituting
the expressions we found for the $p_i^{(D,E)}$ the
result for the mean value is:
$$\langle x_1(i)\rangle=
\int_{[0,1]^L}d^Lx~ x_1 p_i
={1\over{2L}}+{1\over2}\cdot{{L-1}\over{L+i}}$$
which, apart from the term $1/(2L)$ coming from the contribution
of the value replaced at each update, goes to zero as $i^{-1}$.

\section{Master equation for the n.n. model}

We move to the derivation of the master equation in the case of a simplified
version of the n.n. model, where besides the maximum element only one of the
two adjacent ones is updated, chosen with equal probability to preserve
left-right symmetry.
The partitioning into regions $I_{A,B}$ is not anymore sufficient,
in the sense that at each update the regions of analyticity of the probability
distribution break up into finer and finer regions.
Then we have to decompose into domains $I_{\vec{A}}$ where for brevity we use the
vector notation $\vec{A}=(A_1,\dots,A_L)$. To this vector we associate the permutation
$\sigma_A(k)=A_k$. Let's introduce some notations, operations on vectors $\vec{A}$ and
rules. If we take away the element $A_k$ from the vector $\vec{A}$ the resulting
$L-1$ dimensional vector is denoted as
$$\vec{A}(A_k)=(A_1,\dots,A_{k-1},A_{k+1},\dots,A_L)$$
and if we take away element $T$ and move it in the $k$ position:
$$\vec{A}(T:k)=(\vec{A}(T)_1,\dots,\vec{A}(T)_{k-1},T,\vec{A}(T)_k,
\vec{A}(T)_{k+1},\dots).$$
Some useful identities are:
$$\Delta_T\Delta_{\vec{A}}=\delta_{T,A_1}\Delta_{\vec{A}(T:1)}$$
$$\Delta_{\vec{B}}\Delta^T_{\vec{A}(T)}=\delta_{\vec{B}(T),\vec{A}(T)}
\Delta_{\vec{B}}$$
$$\delta_{\vec{A}(T),\vec{B}(T)}F(\vec{B})
=\delta_{\vec{A}(T),\vec{B}(T)}F\left(\vec{A}(T:\sigma^{-1}_B(T))\right)$$
where $F(\vec{B})$ is any expression involving vector $\vec{B}.$ 
Let's start by rewriting the result of the 
removal of the maximum element in terms of the finest decomposition that is 
$p^{\vec{B}}(\vec{x};T)$ in terms of $p^{\vec{B}'}(\vec{x})$.
Given that:
$$p(\vec{x}|I_T)=L\Delta_T(\vec{x})p(\vec{x})
=L\sum_{\vec{A}(T)}\Delta_{\vec{A}(T:1)}p^{\vec{A}(T:1)}$$
we have
$$p(\vec{x};T)
=\int dx_T~p(\vec{x}|I_T)
=L\sum_{\vec{A}(T)}\Delta^T_{\vec{A}(T)}\int^1_{x_{\vec{A}(T)_1}}dx_T~
p^{\vec{A}(T:1)}$$
and then
$$p^{\vec{B}}(\vec{x};T)
=L\sum_{\vec{A}(T)}\delta_{\vec{B}(T),\vec{A}(T)}
\int^1_{x_{\vec{A}(T)_1}}dx_T~p^{\vec{A}(T:1)}
=L\int^1_{x_{\vec{B}(T)_1}}dx_T~p^{\vec{B}(T:1)}.$$ 
The next step is to find the change in the $p(\vec{x})$ after having removed 
an arbitrary element $D$. The new probability distribution 
is denoted $p(\vec{x};[D])$:
$$p(\vec{x};[D])
=\sum_{\vec{B}}\int dx_{D}~\Delta_{\vec{B}}~p^{\vec{B}}(\vec{x})
=\sum_{\vec{B}}\Delta^{D}_{\vec{B}(D)}
\int^{x_{\pi_B^{-1}(D)}}_{x_{\pi_B(D)}}dx_{D}~p^{\vec{B}}(\vec{x})$$
where the operator $\pi_B^n$ on an element is given by $\pi^n_B(B_k)=B_{k+n}$ 
and by definition $x_{\pi^{-1}_B(B_1)}\equiv1$,
$x_{\pi^1_B(B_L)}\equiv0$ $\forall\vec{B}$.
Projecting
$$p^{\vec{A}}(\vec{x};[D])
=\sum_{\vec{B}}\delta_{\vec{A}(D),\vec{B}(D)}
\int^{x_{\pi_B^{-1}(D)}}_{x_{\pi_B(D)}}dx_{D}~p^{\vec{B}}(\vec{x})$$
$$=\sum_{\vec{B}}\delta_{\vec{A}(D),\vec{B}(D)}
\left[\int\dots\right]_{\vec{B}\rightarrow\vec{A}(D:\sigma^{-1}_B(D))}
=\sum_k\int^{x_{\pi_{\vec{A}(D:k)}^{-1}(D)}}_{x_{\pi_{\vec{A}(D:k)}(D)}}
dx_{D}~p^{\vec{A}(D:k)}(\vec{x})$$
$$=\sum_k\int_{x_{\vec{A}(D)_k}}^{x_{\vec{A}(D)_{k-1}}}dx_{D}~
p^{\vec{A}(D:k)}(\vec{x}).$$
Denoting the process of updating successively the maximum element and another one
by $p^{\vec{A}}(\vec{x};T,[D])$ we have for the corresponding master equation:
$$p_{i+1}^{\vec{A}}(\vec{x})=\hat{\sum}_{T,D}~\mu_i(T,[D])
p_i^{\vec{A}}(\vec{x};T,[D])$$
\begin{equation}
=\alpha\hat{\sum}_{T,D}\sum_k
\int_{x_{\vec{A}(D)_k}}^{x_{\vec{A}(D)_{k-1}}}dx_{D}~
\int^1_{x_{\vec{A}(D:k)(T)_1}}dx_T~p_i^{\vec{A}(D:k)(T:1)}
\label{equ:20}\end{equation}
where in the case of the n.n. model the hat indicates sum over n.n., 
$\mu_i(T,[D])=(2L)^{-1}$ and $\alpha=1/2$.
This is a mapping for the set of $L!$ functions $p^{\vec{A}}(x_1,\dots,x_L)$. 
In the case of a system on a ring, due to rotational and
reflection symmetries, there are only $(L-1)!/2$ independent functions. 
In fact if $\vec{A}$ and $\vec{B}$ are related by a translation 
(a cyclic permutation) and/or a reflection we have:
$p^{\vec{A}}(\vec{x})=p^{\vec{B}}(\vec{y})$ with
$x_{A_k}=y_{B_k}.$ 
The number of functions in
the game can be further reduced, as we'll see in the next section.

\section{the case L=4}

To test the correctness of the last master equation we'll apply it  
to the case $L=4$ which is the minimum size able to generate 
information correlations.
We choose as independent the functions $p^{(1,2,3,4)}$, $p^{(1,2,4,3)}$ 
and $p^{(1,3,2,4)}$. Moreover, these three functions can be
expressed in terms of a single two variables function, $g(x,y)$, in the
following way:
$$p^{(1,2,3,4)}(x_1,x_2,x_3,x_4)=[g(x_1,x_2)+g(x_2,x_3)+g(x_3,x_4)+g(x_1,x_4)]/4$$
$$p^{(1,2,4,3)}(x_1,x_2,x_3,x_4)=[g(x_1,x_2)+g(x_2,x_3)+g(x_4,x_3)+g(x_1,x_4)]/4$$
$$p^{(1,3,2,4)}(x_1,x_2,x_3,x_4)=[g(x_1,x_2)+g(x_3,x_2)+g(x_3,x_4)+g(x_1,x_4)]/4.$$
In terms of $g$ the master equation (\ref{equ:20}) becomes:
$$g'(x,y)={1\over2}\left[(1-x^2)g(x,y)+g_x(x,x)+xg_x(x,y)+(1-x)g_x(y,y)+
\right.$$ $$\left.
(1-x)g_y(x)+(1-x)g_y(y)+2g_{xy}(x)+2G(x)\right]$$
where we have defined
$$g_x(x,y)=\int_x^1 d\xi~g(\xi,y)$$
$$g_y(x)=\int_0^x d\eta~g(x,\eta)$$
$$g_{xy}(x)=\int_0^x d\eta~g_x(x,\eta)$$
$$G(x)=\int_x^1 d\xi~g_x(\xi,\xi).$$
The two variables function $g$ can be directly interpreted as a 
probability distribution: $g(x_A,x_B)$ is the probability distribution of
the values of elements $A$ and $B$ with the condition $x_A>x_B$ after the update of the
other two. This allows us to express the one-element
probability distribution (without the trivial contribution
coming from the two updated elements) as:
$$f(x)=\int_0^x d\eta~g(x,\eta)+\int_x^1 d\xi~g(\xi,x)=g_y(x)+g_x(x,x).$$
Starting with the uniform initial condition, $g_0(a,b)=1$, the master equation can be 
iterated exactly. 
The resulting mean value $m_i=\int_0^1 dx~xf_i(x)$ is
in perfect agreement with simulations.

\section{conclusions}

Our main point was to show the presence and the origin of correlations related to
the peculiar mechanism of extremal dynamics 
and to develop a formalism to treat them exactly in the case of Bak-Sneppen models. 
This leads in general to a linear mapping of
$L!$ functions in $L$ variables, the probability distributions of the element values
subjected to one of the $L!$ possible orderings. This mapping can be further simplified 
by taking into consideration the symmetries of the dynamics as it was shown in the case
of a four elements system where the $24$ functions in $4$ variables are reduced to only 
one in two variables. The hope is to reduce the generic finite size system 
to something tractable, at least as a starting point for approximations.

\section*{acknowledgments}
The author thanks M. Droz and P. Cedraschi for many usefull discussions.
This work was supported by the Swiss National Science Foundation.




\begin{references}
\bibitem{BakSne}P. Bak and K. Sneppen, Phys. Rev. Lett. {\bf 71} 4083 (1993)
\bibitem{Mar1}M. Marsili, J. Stat. Phys. {\bf 77} 733 (1994)
\bibitem{Mar2}M. Marsili, Europhys Lett. {\bf 28} 385 (1994)
\bibitem{deBDerFlyJacWet}J. de Boer, B. Derrida, H. Flyvbjerg, A.D. Jackson 
and T. Wettig, Phys. Rev. Lett. {\bf 73} 906 (1994)
\bibitem{deBJacWet}J. de Boer, A.D. Jackson and T. Wettig, Phys. Rev. E 
{\bf 51} 1059 (1995)
\bibitem{PacMasBak}M. Paczuski, S. Maslov and P. Bak, Phys. Rev. E {\bf 53}
414 (1996)
\bibitem{GabCafMarPie}A. Gabrielli, R. Cafiero, M. Marsili and L. Pietronero, 
cond-mat/9702176 
\end{references}
\end{document}